\documentclass{article}

\usepackage{arxiv}

\usepackage[utf8]{inputenc} 
\usepackage[T1]{fontenc}    
\usepackage{hyperref}       
\usepackage{url}            
\usepackage{booktabs}       
\usepackage{amsfonts}       
\usepackage{nicefrac}       
\usepackage{microtype}      
\usepackage{lipsum}

\title{Spatio-temporal small area surveillance of the Covid-19
pandemics}

\author{
    Miguel A. Martinez-Beneito
   \\
    Department of Statistics and Operations Research. \\
    University of Valencia. \\
  Burjassot (Valencia). \\
  \texttt{\href{mailto:miguel.a.martinez@uv.es}{\nolinkurl{miguel.a.martinez@uv.es}}} \\
   \And
    Jorge Mateu
   \\
    Department of Mathematics \\
    University Jaume I of Castellon \\
  Castelló de la Plana \\
  \texttt{\href{mailto:mateu@uji.es}{\nolinkurl{mateu@uji.es}}} \\
   \And
    Paloma Botella-Rocamora
   \\
    Subdirecció General d'Epidemiologia, Vigilància de la Salut i
Sanitat Ambiental. \\
    Conselleria de Sanitat Universal i Salut Pública. Generalitat
Valenciana. \\
  Valencia. \\
  \texttt{\href{mailto:botella_pal@gva.es}{\nolinkurl{botella\_pal@gva.es}}} \\
  }

\newlength{\cslhangindent}
\newenvironment{cslreferences}%
  {\setlength{\parindent}{0pt}%
  \everypar{\setlength{\hangindent}{\cslhangindent}}\ignorespaces}%
  {\par}

\begin{document}
\maketitle

\def\tightlist{}

\begin{abstract}
The emergence of Covid-19 requires new effective tools for
epidemiological surveillance. Spatio-temporal disease mapping models,
which allow dealing with highly disaggregated spatial and temporal units
of analysis, are a priority in this sense. Spatio-temporal models
provide a geographically detailed and temporally updated overview of the
current state of the pandemics, making public health interventions to be
more effective. Moreover, the use of spatio-temporal disease mapping
models in the new Covid-19 epidemic context, facilitates estimating
newly demanded epidemiological indicators, such as the instantaneous
reproduction number \(R_t\), even for small areas. This, in turn, allows
to adapt traditional disease mapping models to these new circumstancies
and make their results more useful in this particular context. In this
paper we propose a new spatio-temporal disease mapping model,
particularly suited to Covid-19 surveillance. As an additional result,
we derive instantaneous reproduction number estimates for small areas,
enabling monitoring this parameter with a high spatial disaggregation.
We illustrate the use of our proposal with the separate study of the
disease pandemics in two Spanish regions. As a result, we illustrate how
touristic flows could haved shaped the spatial distribution of the
disease. In these real studies, we also propose new surveillance tools
that can be used by regional public health services to make a more
efficient use of their resources.
\end{abstract}

\keywords{
    Covid-19
   \and
    disease mapping
   \and
    instantaneous reproduction number
   \and
    spatio-temporal modeling
  }

\hypertarget{introduction}{%
\section{Introduction}\label{introduction}}

Covid-19 has emerged as a novel health problem, posing new challenges
and pressure to health systems, in particular public health services. As
a consequence, new tools and research are required in order to solve the
problems arising from this pandemic. On one hand, biomedical research is
required in order to either develop new vaccines that could stop, or at
least decrease, the spread of the disease or new treatments that could
help to handle the disease in clinic terms. On the other hand, new
public health methodologies and tools are also required in order to
address the corresponding health threaten that affects population
worldwide. Indeed, it has been recently stated that ``countries should
not ease restrictions until they have robust systems in place to closely
monitor the infection situation'' (Han et al. 2020). In this context,
the development of effective epidemiological surveillance tools that
could make it possible to focus public health efforts on particular
locations and moments are a high priority. These tools would be a
valuable help in the current setting, where the public health resources
demanded by the pandemic are frequently higher than those available;
thus, having a precise guidance on where and when actions should be
taken to control the pandemic is crucial.

A compulsory requirement of effective epidemiological surveillance
systems is dealing with sufficiently disaggregated data, both in space
and time. Surveillance systems working with either large spatial units
or long time units will be neither specific nor updated enough as to
provide useful results in epidemiological terms. Anyway, handling highly
disaggregated data in terms of either space or time poses statistical
challenges, known in the statistical literature as small areas
estimation problems (Rao 2003), that should be solved if results with
sufficient reliance are really wanted. In this sense, the small area and
disease mapping literature (Lawson 2018; Martinez-Beneito and
Botella-Rocamora 2019) could be useful for this purpose, and the use of
spatio-temporal disease mapping models could be of particular interest.

Most epidemiological surveillance systems use incidence rates as main
indicator of the current state of any disease outbreak. A vast
literature exists on the small areas smoothing of rates, in particular
standardised rates, which allow visualising current incidence levels,
even dealing with highly disaggregated areal data (Besag, York, and
Mollié 1991; Leroux, Lei, and Breslow 1999). Nevertheless, Covid-19
pandemics has focused substantial attention also on the instantaneous
reproduction number \(R_t\) (Nishiura and Chowell 2009), but no small
areas estimation literature is available for this indicator. For each
time period \(t\), \(R_t\) can be interpreted as the basic reproduction
number of the disease, \(R_0\) (Milligan and Barrett 2014),
corresponding to that particular time period. Thus, \(R_t\) could be
interpreted as a temporally dynamic version of \(R_0\). The basic
reproduction number \(R_0\) is defined as the number of people that
would be infected, in average, in a completely susceptible population by
a single infected person. For \(R_0>1\) the pandemics has an explosive
increasing evolution while for \(R_0<1\) it tends to fade out in a given
period of time. As made clear in this Covid-19 pandemics, where
mitigation actions have been undertaken at different time moments, there
is no a single static value for \(R_0\) inherent to the disease
(Wilasang et al. 2020; Fang, Nie, and Penny 2020). On the contrary,
\(R_t\) shows substantial temporal variability and this statistic is an
interesting quantity that allows monitoring the rate of spread of the
disease at every single moment of the period of study, addressing if the
epidemiological control measures undertaken are enough, or not, to stop
the spread of the disease. While temporal variability of \(R_t\) is
frequently assessed by surveillance systems, the geographical
variability of this indicator is usually unknown, at least with a
sufficiently high disaggregation level. This avoids that specific
epidemiological measures are taken wherever they are required according
to this criterion.

Estimating \(R_t\) can be problematic in some particular settings. For
example, the scarce daily information on the observed counts analysed by
Cori et al. (2013) made them to consider weekly temporal units of
analysis, where the cases observed during the week before day \(t\) were
gathered in order to estimate \(R_t\) on that precise day. As a
consequence, they considered consecutive overlapping time windows for
their analysis (one for each \(R_t\)), ignoring the dependence that this
could yield on the data. Ignoring such dependence could have an impact
on some issues, as for example the width of confidence intervals for the
different \(R_t\)'s. Moreover, the weekly character of those time
windows makes the observed cases of the disease to be averaged during
the last observed week, making \(R_t\) less sensible to the latest
incidence changes that the time series could show. This makes the
interest of \(R_t\) decrease as an epidemiological surveillance tool.
This modelling approach is implemented in the \texttt{EpiEstim} (Cori
2020) \texttt{R} package, which is being extensively used for many
public health institutions for Covid-19 surveillance tasks (see Tebé et
al. (2020) or Centro Nacional de Epidemiología. Insituto de Salud Carlos
III (2020) for just two examples).

The statistical problems of \(R_t\) estimation just mentioned are mainly
due to the few observed cases available per time unit when the units of
analysis are small. However, these problems would be exacerbated if
several spatial units of study were considered, instead of a single
unit, since those units would split the observed daily cases into
smaller quantities. In that case, a small areas modelling framework
would be required in order to yield sensible incidence estimates sharing
information between neighbouring sites, since otherwise those estimates
would mainly reproduce the high amount of noise that such small figures
frequently show.

In this paper we propose a small areas spatio-temporal model
specifically devised for Covid-19 incidence data. With our proposal we
seek to develop advanced surveillance tools that allow monitoring the
disease at high disaggregation levels. In real terms, this would make it
possible to implement specific epidemiological control measures wherever
and when they are required according to detailed and specific
information. Additionally, our spatio-temporal model also allows to
visualize the temporal evolution of the geographical pattern of the
disease. Visualising that evolution helps understanding the performance
of the disease and shows how the different control measures implemented
throughout the period of study have modified that spatial distribution.

In addition, we apply the spatio-temporal model developed in this paper
to two real Covid-19 data sets. These data sets correspond to
surveillance data of two Spanish regions with a high spatial and
temporal disaggregation. Specifically, these data sets contain the daily
observed incident cases for a detailed division of both regions, with
less than 10000 people in average per spatial units in both cases, and
with substantially lower population in many particular cases. The
application of our model to these data sets illustrate its impact in
practical terms.

This paper is organised as follows. Section 2 introduces a general
spatio-temporal modelling proposal for the analysis of Covid-19
surveillance data sets of potential use in many different regions.
Section 3 shows the application of that model to the analysis of the
Covid-19 pandemics in two Spanish regions. This section illustrates the
real use in practice of our strategy and the corresponding results.
Finally, Section 4 discusses some aspects of the model introduced and
its use for Covid-19 surveillance.

\hypertarget{modelling-proposal}{%
\section{Modelling proposal}\label{modelling-proposal}}

We consider the following general setting that, in principle, could be
easily adapted or directly used for many small areas Covid-19
surveillance studies. Let us assume that we have a region of study
divided into \(I\) spatial units, of small size in a statistical sense,
observed during \(J\) consecutive days. Let \(O_{ij}\) denote the
observed number of incident Covid-19 cases for the \(i\)-th spatial unit
\((i=1,\ldots,I)\) and the \(j\)-th day \((j=1,\ldots,J)\) of the period
of study. We consider the observed cases to be independently Poisson
distributed, given the underlying process \(\lambda\), as
\(O_{ij}\sim Poisson(\lambda_{ij})\), and model the underlying process
as
\[log(\lambda_{ij})=log(P_i)+\boldsymbol{\gamma}_{DoW(j)}+(\boldsymbol{\beta}\mathbf{X})_{ij}+\epsilon_{ij}\]
where:

\begin{itemize}
\tightlist
\item
  \(P_i\) is the population living at the \(i\)-th spatial unit during
  the period of study. This quantity is considered constant in time
  during all that period.
\item
  \(\boldsymbol{\gamma}\) is a vector, of length 7, intended to model
  the supposedly important cyclic effect of the different week days on
  the incidence rates. The term \(DoW(j)\) denotes a function that
  returns the day of the week, as a value between 1 and 7 corresponding
  to the \(j\)-th day of the period of study.
\item
  \(\mathbf{X}\) is a \(K\times J\) design matrix, where each column
  corresponds to a different day and each row corresponds to a different
  basis function, intended to model the incidence rates time trend for
  the different spatial units. \(X_{kj}\) stands for the value of the
  \(k\)-th basis function on the \(j\)-th day. This matrix is fixed by
  design and no inference is made on \(\mathbf{X}\).
\item
  \(\boldsymbol{\beta}\) is an \(I \times K\) matrix with the
  coefficients of the basis functions used for modelling the incidence
  rate time trend for each spatial unit. The variability of the
  different rows of \(\boldsymbol{\beta}\) allow the incidence rates of
  the spatial units to follow different trajectories.
\item
  \(\boldsymbol{\epsilon}\) is an \(I \times J\) matrix of independent
  random effects. This unstructured term is intended to model the
  overdispersion of the Poisson process beyond the variability induced
  by the rest of terms in the model. Covid-19 incident cases tend to
  cluster in some particular days for each spatial unit due to either
  the outbreak nature of these cases that makes them to be grouped, or
  some other artifacts, such as the sampling of several suspicious cases
  around the first index case of an outbreak, which will usually be all
  tested in a same day. These artifacts may produce abnormally high
  numbers of cases for some particular days and locations, beyond the
  smooth time trend of the disease, that we try to take into account
  with this term.
\end{itemize}

The spatio-temporal variability of the process is mainly modelled by the
product \(\boldsymbol{\beta}\mathbf{X}\). This product relies on an
appropriate basis of functions suitable for modelling the temporal
variability of the disease. Once these functions are chosen, the
coefficients \(\boldsymbol{\beta}\) combine them in order to define the
time trends for the disease; a different time trend for each spatial
unit. Moreover, the number of functions in the basis \(K\) may be
different to the number of days \(J\) in the period of study, in
particular it could be substantially lower. This makes the time trend
for each spatial unit depend on just a few parameters \(K\), much lower
than the number of observations on that spatial unit \(J\), making a
low-dimensional fit of that time trend, quite convenient in
computational terms.

Although we could formulate the model above for a general basis of
functions \(\mathbf{X}\), we find it particularly convenient the use of
splines as basis of functions for modelling the temporal variability of
the incidence rates. In particular, we propose to use a basis of natural
cubic B-splines for that goal. Besides its computational convenience,
B-splines have the advantage of having compact support, thus the
elements of this basis model the temporally local performance of the
incidence curve during an interval of days, making the interpretation of
the corresponding coefficients \(\boldsymbol{\beta}\) quite
straightforward. Moreover, the use of natural splines makes the fit of
the incidence trends to be particularly sensible at the extremes of the
period of study, specifically at the end of that period. This is
particularly convenient for surveillance purposes, where attention is
usually put on these precise days of the period of study.

Our proposal models the incidence time trends by means of the fitted
splines. On the other hand the spatial smoothing of the incidence rates
is carried out through the matrix of spline coefficients
\(\boldsymbol{\beta}\), in particular the spatial dependence of its
cells. Specifically, we model those cells as
\(\beta_{ik}=\mu_k+\beta_{ik}^*\), where \(\mu_k\) models the mean
value, across the set of spatial units, of the coefficient corresponding
to the \(k\)-th spline in the basis. These values will change according
to the mean incidence level at each moment of the period of study, and
the vector \(\boldsymbol{\mu}=(\mu_1,\ldots,\mu_K)\) accounts for those
overall temporal variations. On the other hand, the matrix
\(\boldsymbol{\beta}^*\) models the spatial variability of the different
incidence time trends. This term allows each spatial unit to follow a
different time trend. For inducing a smooth spatial performance, we
model the columns of this matrix as spatial random effects following
some Gaussian Markov random field (Rue and Held 2005). In this sense we
could use for example a proper CAR distribution, a combination of an
Intrinsic CAR and an heterogeneous random effect, as proposed by Besag,
York, and Mollié (1991) or the CAR distribution of Leroux, Lei, and
Breslow (1999). In our case we opt by using the latter option for
inducing spatial dependence on the incidence rates, i.e., we assume the
following set of conditional distributions
\[\beta^*_{ik}\sim N\left(\frac{\rho_k}{1-\rho_k+\rho_k n_i}\sum_{i'\sim i}\beta^*_{i'k},\frac{(\sigma_{\beta}^2)_k}{1-\rho_k+\rho_k n_i}\right).\]
where the subindex \(i'\sim i\) denotes all spatial units \(i'\)
adjacent to spatial unit \(i\) and \(n_i\) denotes the number of
neighbours of that area. In these conditional distributions \(\rho_k\)
takes values in the interval \([0,1]\), and this parameter may reproduce
either an heterogeneous process for \(\rho_k=0\), an intrinsic CAR
distribution for \(\rho_k=1\) or intermediate process, with stronger or
weaker spatial dependence, for \(\rho_k \in ]0,1[\). Considering
different \(\rho_k\) and \((\sigma_{\beta})_k\) parameters for each
element in the basis of functions allows reproducing different spatial
strengths and different spatial variabilities, respectively, for the
different moments of the period of study. As a consequence, this
reproduces a non-separable spatio-temporal correlation structure
(Torres-Avilés and Martinez-Beneito 2015). These assumptions seem
reasonable when the pandemic could show very different spatial features
at the different phases of the pandemic waves or even at the different
waves of the pandemic.

The term \(\boldsymbol{\beta}\mathbf{X}\) in the linear predictor
models, for each spatial unit, the temporal variability of the Covid-19
incidence along the period of study. On the other hand, the term
\(\boldsymbol{\gamma}\) is in charge of modelling the cyclic variability
within weeks. This term is mainly required to take into account the
different incidence rates that weekends could have in comparison to
weekdays. We will consider the \(DoW\) function to be equal to 1 for the
first day of study and it will increase, day by day, until 7.
Afterwards, \(DoW\) will reproduce this cycle repeatedly. The first
element of the vector \(\boldsymbol{\gamma}\), \(\gamma_1\), will be
fixed to 0 and therefore the rest of elements of \(\boldsymbol{\gamma}\)
will model the departures of the incidence rates of the rest of week
days, in comparison to the first. Improper uniform prior distributions
on the whole real line are assigned to \(\gamma_2,\ldots,\gamma_7\).

Finally, regarding the cells of the matrix of random effects
\(\boldsymbol{\epsilon}\), they are modelled as independent
\(N(0,\sigma^2_{\epsilon})\) variables. In this case,
\(\sigma^2_{\epsilon}\) is also estimated within the model and a
\(Uniform(0,c)\) distribution, for a large enough (non informative)
value \(c\), is assigned to the standard deviation
\(\sigma_{\epsilon}\).

\hypertarget{small-areas-estimation-of-the-instantaneous-reproduction-number-r_t}{%
\subsection{\texorpdfstring{Small areas estimation of the instantaneous
reproduction number
\(R_t\)}{Small areas estimation of the instantaneous reproduction number R\_t}}\label{small-areas-estimation-of-the-instantaneous-reproduction-number-r_t}}

One advantage of our proposal for spatio-temporal modelling of the
COVID-19 incidence, is that we could provide small areas estimates of
some epidemiological quantities of particular interest, such as the
instantaneous reproduction number \(R_t\) for all days \(t\) of the
period of study, and for the small areas division of the region of
study. Specifically, we will estimate \(R_{it}\), the instantaneous
reproduction number for each (small) spatial unit \(i\) and each day
\(t\) of the period of study.

In a non-spatial setting, leaving aside the influence of the prior
distribution used for the different \(R_t\)'s, the widely used
\texttt{EpiEstim} package of \texttt{R} estimates the \(R_t\)'s as
\(O_t/(\sum_{s=1}^S O_{t-s}w_s)\), where \(O_t\) denotes the incident
cases for the whole region at the \(t\)-th day of study. In this
expression \(w_s\) denotes the \emph{infectivity function} that
quantifies, for each lag time \(s\), the probability of observing an
\(s\)-days difference between two cases, one primary and one secondary
case of the disease, being \(S\) an upper bound for the subindex \(s\).
The incident cases \(O_t\) could be quite low for some days of the
period of analysis, even for the whole region of study, what could make
\(R_t\) to be noisy and unreliable. As a potential solution, these
authors propose to use overlapping time windows of size \(\tau\) instead
of single days and therefore estimate \(R_t\) instead as
\[\left(\sum_{k=t-\tau+1}^t O_k\right)/\left(\sum_{k=t-\tau+1}^t \sum_{s=1}^S O_{k-s}w_s\right).\]
This illustrates how the \(R_t\) estimation poses particular problems,
mainly when small counts are frequent. This will be the usual case when
working with small sized spatial units, justifying the use of
particularly suited methods if these indicators were to be calculated on
that particular setting. According to the disease mapping literature, if
\(O_i \sim Poisson(\lambda_i)\) are the observed counts of some disease
at the \(i\)-th spatial unit, and \(E_i\) the corresponding expected
counts, the Standardised Incidence Ratio (SIR) for that spatial unit is
defined as \(SIR_i=O_i/E_i\). This risk estimate shows similar small
areas problems to those just introduced, arising from the small values
that \(O_i\) and \(E_i\) usually take. These problems are normally
solved by considering an smoothed version of those SIRs defined as
\(SIR_i=\lambda_i/E_i\), where \(\lambda_i\) is estimated taking into
account the incident cases of that spatial unit and the underlying
hypotheses of the model, that will usually induce spatial dependence on
\(\lambda_i\). We follow a similar approach for deriving smoothed
instantaneous reproduction numbers that could solve the small areas
problems of the corresponding unsmoothed indicator.

In parallel to Cori et al. (2013), and following the disease mapping
smoothing ideas just mentioned, we could estimate \(R_{it}\) for the
small area \(i\) and day \(t (=S+1,\ldots,J)\) as:
\(R_{it}=\lambda_{it}/(\sum_{s=1}^S \lambda_{i\,t-s}w_s)\), where
\(\lambda_{it}\) is now the expected incident cases for that spatial
unit and day according to our model. We could set \(\lambda_{it}\) to be
equal to
\(\exp(log(P_i)+\boldsymbol{\gamma}_{DoW(t)}+(\boldsymbol{\beta}\mathbf{X})_{it}+\epsilon_{it})\),
or simply
\(\lambda_{it}=\exp(\boldsymbol{\gamma}_{DoW(t)}+(\boldsymbol{\beta}\mathbf{X})_{it}+\epsilon_{it})\)
since the population term will cancel out in the numerator and
denominator of \(R_{it}\). Nevertheless, we could also remove the cyclic
term in the latest expression, that is
\(\lambda_{it}=\exp((\boldsymbol{\beta}\mathbf{X})_{it}+\epsilon_{it})\)
since, in this manner, we will filter out the cyclic effect that we
could find in \(R_{it}\) as a consequence of that same artifact on the
raw incidence figures and that we will not be typically interested in.
Moreover, if \(\lambda_{it}\) was simply defined as
\(\exp((\boldsymbol{\beta}\mathbf{X})_{it})\), we would obtain a
parsimonious version of \(R_{it}\) where the particular data collecting
features that could be having an influence on \(O_{it}\) are filtered
out and just the smooth spatio-temporal trend captured by the spline
process would be kept. Thus, a parsimonious smoothed instantaneous
reproduction number \(R_{it}\) suitable for small areas could be
obtained as
\[R_{it}=\left( \exp((\boldsymbol{\beta}\mathbf{X})_{it})\right)/\left( \sum_{s=1}^S \exp((\boldsymbol{\beta}\mathbf{X})_{i\;t-s})w_s\right).\]\\

In the case that MCMC inference were performed on the model above, this
would allow to calculate the different smoothed \(R_{it}\)'s at every
step of the MCMC. This would allow to easily draw confidence bands for
the smoothed instantaneous reproduction number for every spatial unit
\(i\). This, in turn, enables to easily draw \(P(R_{it}>1)\) could be
also calculated for each \(i\) and \(t\), and therefore we can assess
the probability of the epidemic to be out of control at every location
and time of the period of study. This quantity, obviously, could be of
evident interest for epidemiologists.

\hypertarget{spatio-temporal-modelling-of-covid-19-for-two-spanish-regions}{%
\section{Spatio-temporal modelling of Covid-19 for two Spanish
regions}\label{spatio-temporal-modelling-of-covid-19-for-two-spanish-regions}}

We now study the Covid-19 pandemics in Spain, in particular in Castilla
y Leon (CyL) and Comunidad Valenciana (CV), 2 out of the 17 Spanish
communities, with 2.3 and 5 million inhabitants, respectively.

The CyL Covid-19 data are publicly available from
\url{https://analisis.datosabiertos.jcyl.es/pages/coronavirus/}. The
period of study for this region covers the interval from March 6th to
October 14th, 2020, which is the period publicly available at that
webpage at October 18th, 2020, the day when this analysis was performed.
Covid-19 data for CyL are published at the health zone level, an
administrative division corresponding to the geographic units covered by
the primary care centers of this region. CyL is divided into 247 health
zones, with population ranging from 441 to 37509 inhabitants. The data
set analysed contains daily observed Covid-19 counts for each of the CyL
health zones. The data set used for this analysis is supplied as
supplementary material to the paper in order to make this part of the
analysis completely reproducible. This data set and the rest of
supplementary material may be found at the webpage:
\url{https://www.uv.es/mamtnez/ETCovid.html}.

CV data have been supplied, under request, by the Health Administration
of this region. As for CyL, the data set contains daily observed
Covid-19 counts for the 542 municipalities that form this region.
Municipalities in CV are towns, in administrative terms, whose
populations range from 17 to 794288 inhabitants in 2019. The median
population per municipality is 1412 inhabitants. In this case the period
of study covers once again the interval from March 6th to October 14th.
The data for this analysis are not supplied since they are not publicly
available.

MCMC Inference was carried out by using \texttt{WinBUGS}. Additionally,
the \texttt{pbugs} \texttt{R} package (Vergara-Hernández and
Martinez-Beneito 2020) was used for running in parallel the 5 MCMC
chains simulated. 2000 MCMC draws were firstly simulated as burn in
period of a total of 5000 draws per chain. Finally, 1 out of each 15
draws were saved yielding therefore a posterior sample of 1000 draws.
Convergence was assessed by means of the Brooks-Gelman-Rubin statistic
and the number of effective simulations (Gelman et al. 2014), which are
directly provided by the \texttt{pbugs} package. Full details on the
implementation of the model, and the rest of results/figures shown
below, can be found at the Rmarkdown and pdf documents in the
supplementary material, which contain the full code used to run this
analysis.

For the long range time trend modelling, we chose a natural B-spline
basis with one node every 2 weeks and 2 final nodes at the borders of
the period of study. Thus, for a total of 223 days of study, the final
basis contained 17 spline functions scattered around all the period of
study, one spline function every 2 weeks. These 17 functions yielded
enough flexibility for modelling the temporal evolution of rates and
\(R_t\)'s, and reduced the computational burden of the model in
comparison to a weekly basis which contained 33 spline functions, and
therefore 33 spatial processes to be estimated within the model.
Regarding the infectivity function \(w_s\), required for calculating the
instantaneous reproduction numbers, we assume a shifted Gamma
distribution of mean 4.7 and standard deviation 2.9. These values have
been estimated for COVID-19 incidence by Nishiura, Linton, and
Akhmetzhanov (2020). Additionally, we have fixed the maximum temporal
lag for calculating \(R_t\) to \(S=25\) since for the mentioned
parameters for \(w_s\) we have \(\sum_{i=1}^{25} w_s>0.9999\), that is,
COVID-19 cases are hardly contagious after those 25 days.

We have run the spatio-temporal model above for these two data sets,
separately. In both occasions we have considered a single spatial
parameter \(\rho\) instead of different parameters \(\rho_k\) for each
element in the basis since we have not found evidence of needing
different spatial parameters. Specifically, we run an additional model
with different \(\rho_k\)'s per basis function and the posterior
confidence intervals of these parameters substantially overlapped. In
our model, with a common \(\rho\), we obtained a posterior mean for
\(\rho\) of 0.51 (95\% credible interval: {[}0.43,0.60{]}) for CyL,
while for CV that posterior mean was 0.45 (95\% credible interval
{[}0.36,0.57{]}), so in both cases the spatial dependence found was
moderate. In contrast, we found evidence of needing different variance
parameters for the spatial processes corresponding to each of the basis
functions. This different variances allow the spatial distribution of
the disease to show varying variability at different moments of the
period of study. Thus, considering different \(\sigma_k\)'s allows the
spatio-temporal process to be temporally heteroscedastic. Specifically,
for CyL the spatial standard deviations for the coefficients of the
splines basis range from 0.88 (95\% credible interval {[}0.65,1.14{]})
to 2.14 ({[}1.76,2.56{]}), while for CV they range from 1.26
({[}0,1.93{]}) to 3.29 ({[}2.76,3.90{]}).

Regarding the overdispersion term \(\boldsymbol{\epsilon}\), its
standard deviation has a posterior mean of 0.56 (95\% credible interval
{[}0.55,0.57{]}) for CyL and of 0.77 ({[}0.75,0.79{]}) for CV. This
shows how the sampling artifacts, which make incident cases to cluster
in some particular days and locations, are stronger in CV than for CyL.
Regarding the cyclic weekly term \(\boldsymbol{\gamma}\), we have found
a difference of 0.94 units between the day with higher and lower
incidence for CyL, and of 1.16 in CV. The 95\% credible intervals for
both days, for both regions, were clearly disjoint and distant, pointing
out the need of this term within the model.

\begin{figure}
\centering
\includegraphics{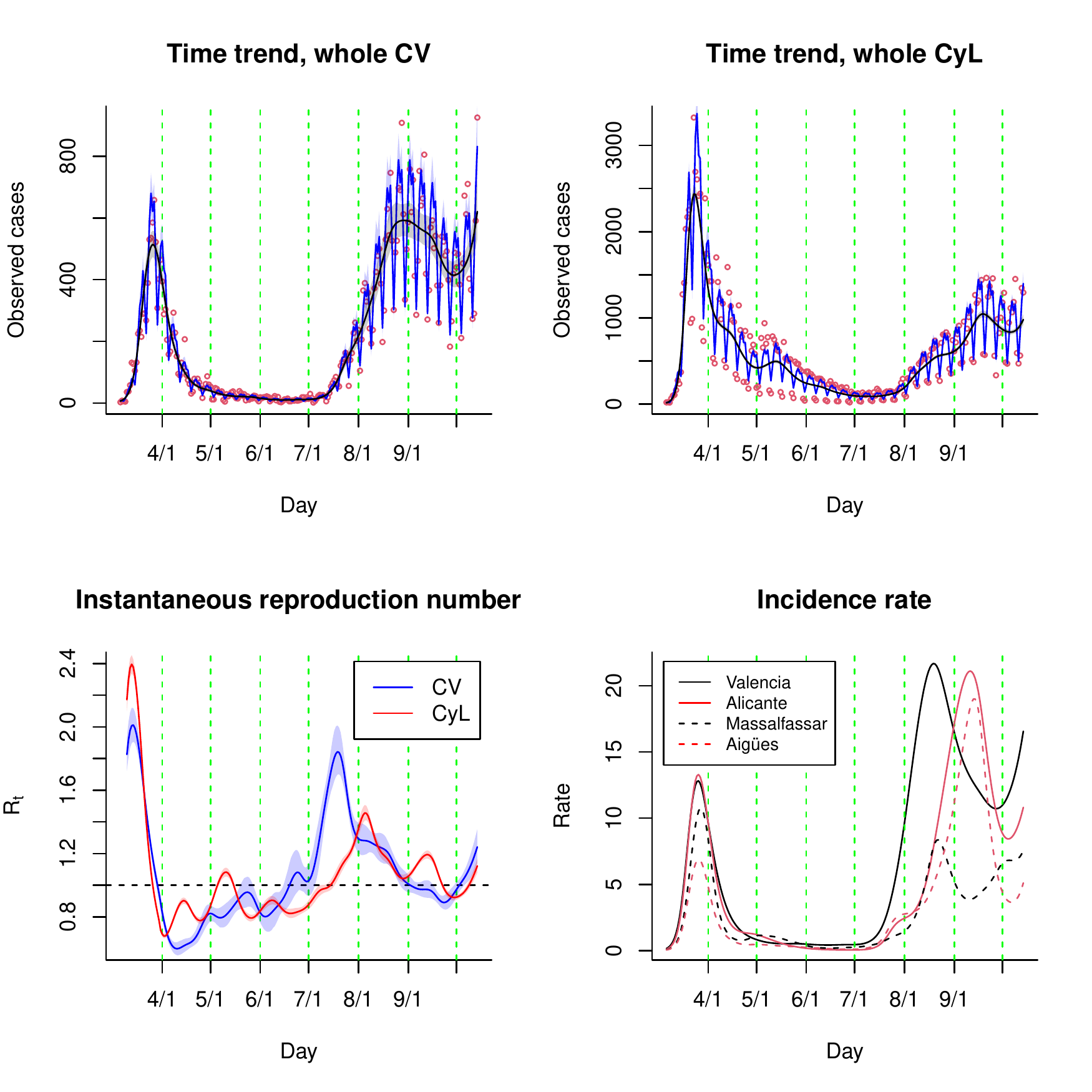}
\caption{Overall time trends for the observed daily cases for both
regions (top row). Estimated \(R_t\) evolution for both regions (bottom
left plot), and daily rates per 100000 inhabitants for 4 municipalities
in CV (bottom right plot).}
\end{figure}

Figure 1 shows several results of the spatio-temporal model for both
regions. The upper row of Figure 1 shows, for both regions, the fitted
mean to the daily observed cases, with the corresponding 95\% credible
bands. We first note the very different scales for both regions, even
though the CyL population is lower than half the CV population. The blue
line in these plots corresponds to the time trends taking into account
both the long range time trend (splines) and cyclic weekly terms. The
effect of the cyclic term is evident. On the other hand, the black line
reproduces just the time trend corresponding to the splines component of
the fit. These curves filter out the cyclic pattern that in general we
will not be interested in to reproduce in the pandemics surveillance.
Both splines curves seem to perfectly capture the time evolution of the
disease, filtering the artifacts that could make the observed cases for
one particular day to deviate below or above that trend. From now on,
the rest of results shown will be based only in the splines component of
the time trend in order to filter out the cyclic effect that we will
typically want to remove.

The bottom left plot of Figure 1 shows the daily evolution of the
instantaneous reproduction number of the pandemics for both regions.
This plot shows the overall time trends, for the whole regions, and
their corresponding 95\% credible bands. Note the effect of the
different amount of observed cases for each region which makes the bands
for CV to be substantially wider than for CyL. The daily \(R_t\)'s in
this plot have been calculated by following the methodology described in
the previous section for each municipality and averaging all those
\(R_t\)'s throughout the whole region. These averages are weighted
according to the different populations of the different spatial units.
These curves have a high epidemiological value since they allow to know
if the pandemics is either increasing (\(R_t>1\)) or decreasing
(\(R_t<1\)) at any moment of the period of study. Moreover, if wanted,
we may also easily assess the probability of any of these two states.
These curves show some evident differences in the pandemics diffusion
for both regions, as for example the different starting times of the
second wave of the pandemics, which started way sooner for CV than for
CyL.

The bottom right plot of Figure 1 shows the rates (per 100000 people)
time trends for 4 CV municipalities. Rates, in contrast to the fitted
observed cases, allow comparisons between regions regardless of their
population sizes. These municipalities correspond to Valencia and
Alicante, the two most populated cities in CV and placed 166 kilometers
away. Additionally, we show also the rates for Massalfassar and Aigues,
the two least populated neighbouring municipalities of Valencia and
Alicante, respectively. As we see in Figure 1, the rates curve for
Massalfassar resembles that of Valencia, in a similar manner to Aigues
and Alicante since, for example, they show concurrent peak epidemics.
This is a consequence of the spatial dependence of the Leroux et al.'s
processes in our model. Anyway, although these curves are similar, we
can see evident departures within both Massalfassar-Valencia and
Aigues-Alicante. For example, the first wave in Aigues showed far lower
rates than for Alicante besides these two cities are neighbours and one
of them has a population more than 300 times higher than the other, what
could make the Alicante time trend to completely determine that of
Aigues. Something similar happens for the second wave of Valencia and
Massalfassar.

\begin{figure}
\centering
\includegraphics{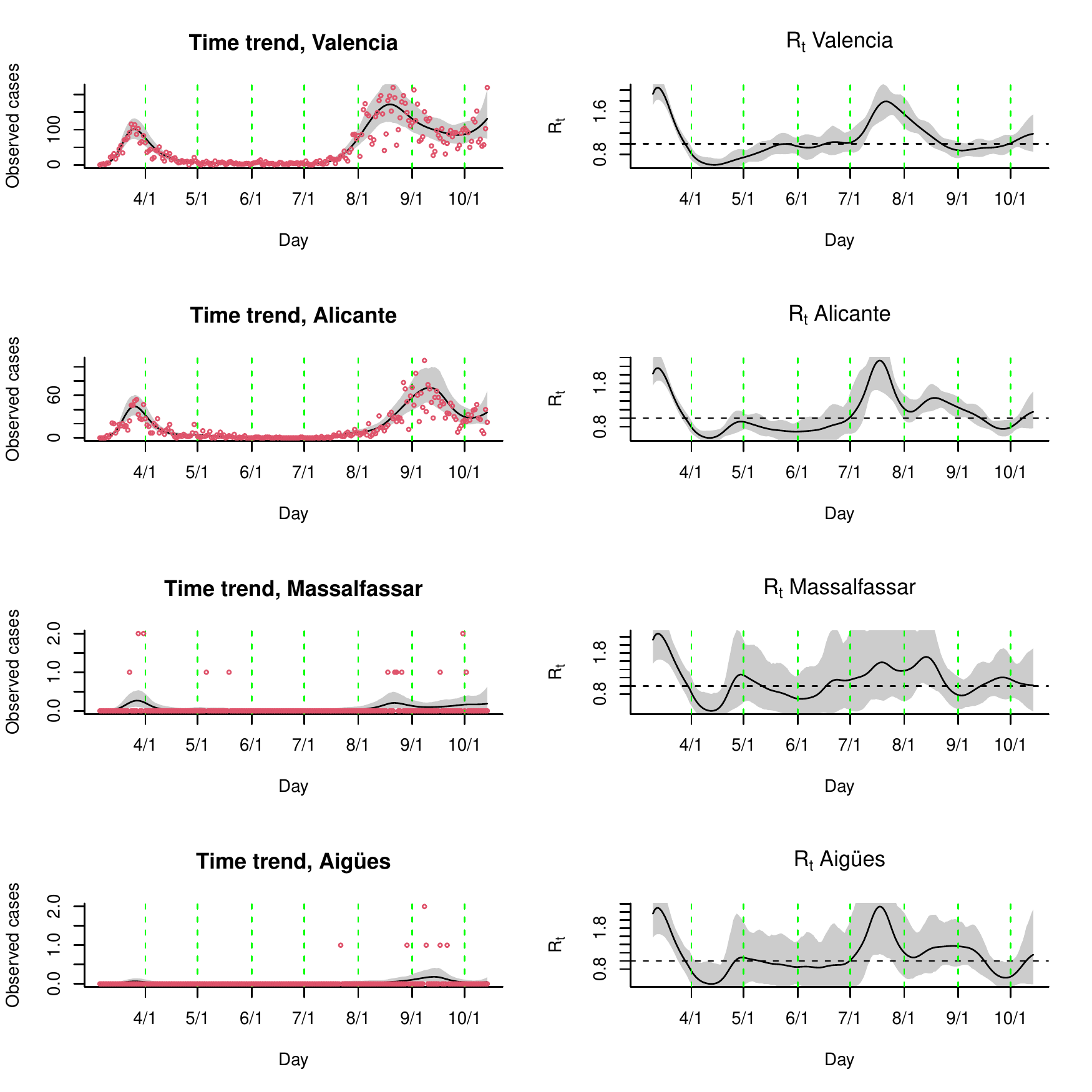}
\caption{Observed daily cases time trend for Valencia, Alicante,
Massalfassar and Aigues (left column), and \(R_t\) time trend for these
same municipalities (right column).}
\end{figure}

Figure 2 shows, once again for Valencia, Alicante, Massalfassar and
Aigues, the fitted mean to the daily observed cases (left-hand-side of
the figure), with the corresponding 95\% credible bands. The
right-hand-side of this figure shows the \(R_t\) temporal evolution for
each of these municipalities. The plots of the observed cases for
Valencia and Alicante show a good fit, thus the biweekly splines used
seem enough for describing the temporal variability of the observed
cases in these large cities. On the other hand, for Massalfassar and
Aigues, the observed cases are quite low due to the small size of these
municipalities. Nevertheless, a time trend is fitted also for these
municipalities which merges the information on the observed cases of
these municipalities and the information provided by their neighbours.
Thus, the time trend for Massalfassar resembles that of Valencia, with a
long right tail at the second wave of the pandemic; larger than that of
Aigues which resembles more the time trend for Alicante. Interestingly,
Aigues does not have any observed case at the first wave of the
pandemics, what explains the discrepancy between Alicante and Aigues at
the first wave of the pandemics already shown in Figure 1.

Regarding the temporal evolution of the \(R_t\)'s for these
municipalities, it is remarkable the longer width of the Massalfassar
and Aigues 95\% credible intervals for \(R_t\) in comparison to those of
Valencia and Alicante. For example, it is rare to find some day during
the second wave of the pandemic in Massalfassar and Aigues where the
corresponding credible interval does not contain the value 1. On the
contrary, for Valencia and Alicante most of the months of July and
August showed \(R_t\) values significantly higher than 1. This is
obviously due to the different populations, and therefore daily observed
cases, of these municipalities. In addition, once again the \(R_t\)
curves of the neighbouring pairs of municipalities resemble each other.
See for example the curves of Alicante and Aigues which describe similar
plateaus and peaks from May to mid September. This is, evidently, a
consequence of the spatial dependence of rates since the observed cases
in Aigues are not strong enough as to describe such a detailed temporal
evolution.

In addition to enhanced (smoothed) spatio-temporal epidemiological
indicators, as the case of the instantaneous reproduction number already
shown, the spatio-temporal model proposed also allows performing deeper
studies and developing surveillance tools of high epidemiological value.
We show now some of these capabilities.

\begin{figure}
\includegraphics[width=1\linewidth]{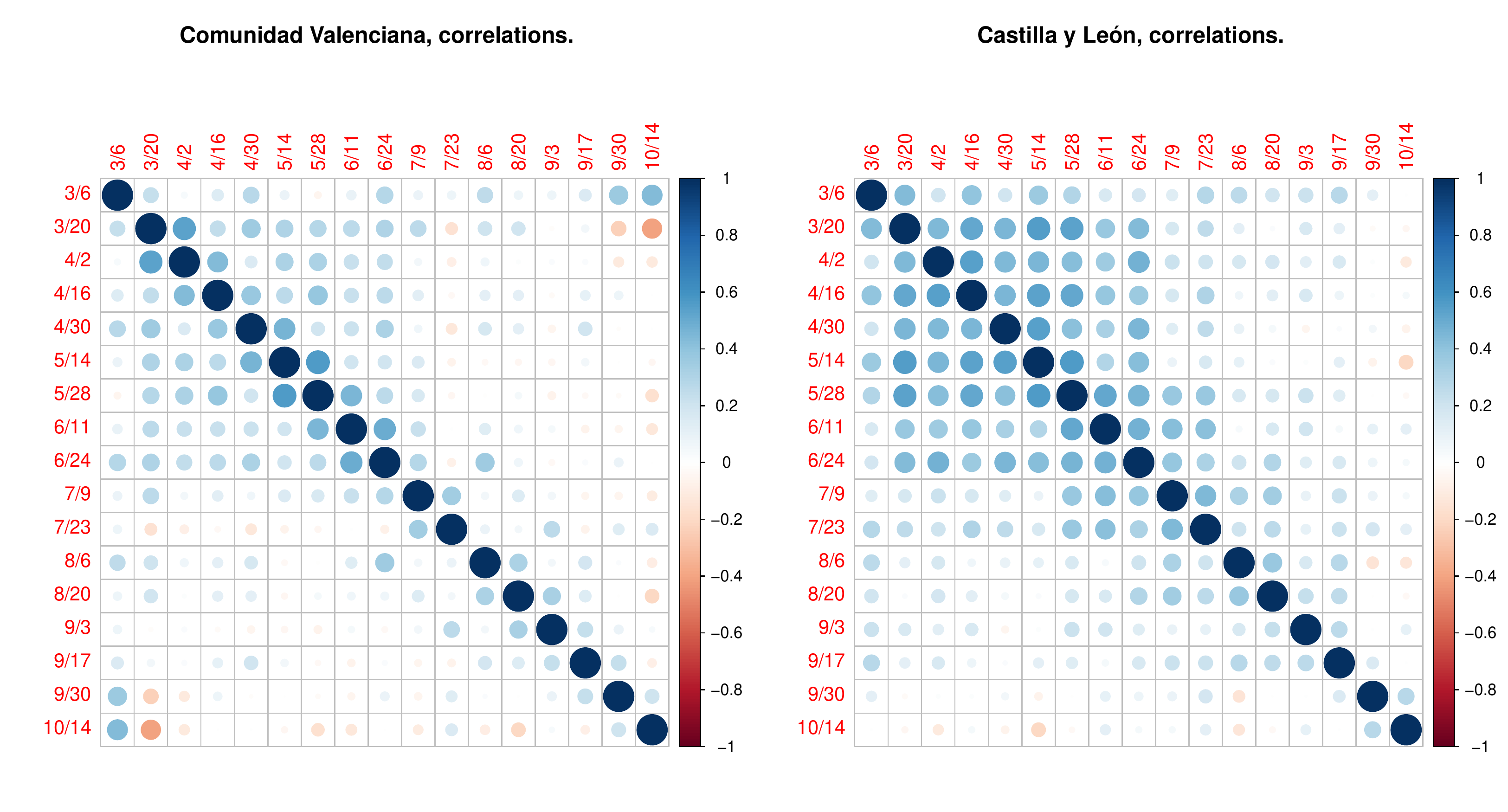} \caption{Correlation plots for the coefficients of the spline basis functions. Names of the rows and columns of the plot correspond to the day showing the peak of the corresponding spline function.}\label{fig:unnamed-chunk-2}
\end{figure}

Figure 3 shows, for each region, a correlation plot for the coefficients
of each function of the spline basis. The names in the rows and columns
of the plot correspond to the day showing the peak of the corresponding
spline function, in order to make temporal sense of the corresponding
coefficients. It can be seen how, for both regions, all the coefficients
previous to July show high correlations, regardless of the moment of the
corresponding peak. On the contrary, this seems to change with the
starting of July. This suggests that the geographical pattern of the
pandemic was steady during the first wave of the pandemics but with the
end of that wave, and also coinciding with the lift of the mobility
restrictions by the Spanish government (active from March 14th to June
21st), the spatial pattern of the disease completely changed. In the
case of CV this change was quite abrupt, the spatial patterns from July
hardly show any correlation with the spatial patterns previous to that
month, and milder in the case of CyL. These different patterns may be
explained by the large touristic sector of CV, whose coastline attracts
millions of tourists each summer. This sector possibly attracted
thousands of tourists coinciding with the starting of the summer and the
lift of restrictions which could make its spatial pattern to change
completely, in contrast to CyL where this change is milder.

Figure 3 shows some additional differences between CV and CyL spatial
pattern evolutions. For example, in CV, we see from July a completely
dynamic evolution of the pandemics since the spatial pattern
corresponding to any spline function of this period hardly shows any
correlation with the spatial pattern of the spline component
corresponding to one month latter. Thus, the pandemics in CV during this
period is moving around with no particular preference for any specific
geographical location of this region, in contrast to the first phase of
the pandemics when population was confined. In contrast, for CyL
correlations show overlapping diagonal squares which points out a less
haphazard evolution of the disease. Evidently, these plots could be
further interpreted with the aid of some additional results. For
example, Figure 5 of the accompanying supplementary material shows how
the first wave of the pandemic in CyL was mainly urban while the second
wave, at least its summer period, showed higher rates in the rural
areas. This points out once again that tourism has possibly taken an
important effect in the second pandemic wave also in this region, since
rural zones receive most of the tourism in this inner zone of Spain,
mainly during the summer period. Questions of this kind are not fully
discussed in this paper for lack of space although they are, evidently,
interesting results that can also be drawn from our model.

Figure 4 also illustrates some of the small areas analysis possibilities
that our spatio-temporal model yields. As illustrated throughout this
paper, both rates and \(R_t\)'s are the main focus of interest of most
Covid-19 epidemiological analyses. These two indicators measure,
respectively, the current state of the pandemics and its temporal
evolution according to the latest data, a kind of derivative of the
incidence curve. Therefore these two quantities are complementary
indicators of the pandemics. Currently, the joint study of these two
indicators has been proposed as a summary measure of the ``risk'' of the
pandemics (\url{https://biocomsc.upc.edu/en/covid-19/Risk\%20Diagrams}).
The joint representation of these two indicators in a single plot has
been proposed and used in an applied epidemiological context, but it has
always been done for large areas (López Codina 2020), due to the
statistical problems for dealing with these indicators when working with
small areas. Nevertheless, our spatio-temporal model could be also used
in this context, making it useful also for small areas problems, as
illustrated in Figure 4.

\begin{figure}
\includegraphics[width=1\linewidth]{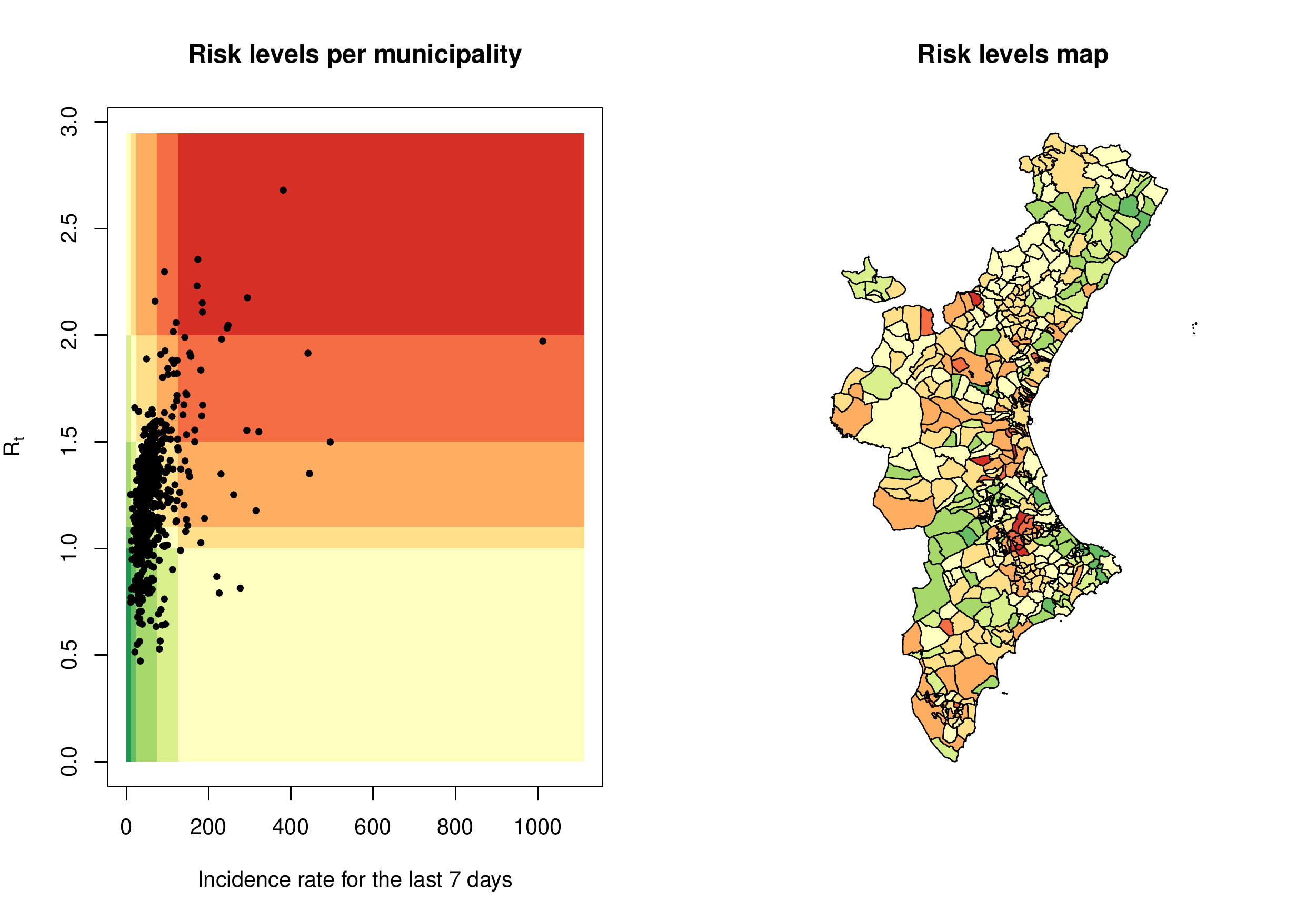} \caption{Risk levels according to smoothed weekly incidence rates and $R_t$'s and its geographical representation.}\label{fig:unnamed-chunk-3}
\end{figure}

The left-hand side of Figure 4 shows a joint plot of the incidence rate
for the last 7 days, jointly with the \(R_t\) for each municipality in
CV at the latest day of the period of study. Results could also be
reproduced for previous days. Incidence rates for the last 7 days have
been defined as 7 times the incidence rate for the latest day of study
in order to represent the most updated information on the disease,
according to the available data. Color cuts in this plot have been set
as the different official risk levels for each of these indicators set
by the Spanish government (Consejo Interterritorial, Sistema Nacional de
Salud 2020), these are: \{10,25,75,125\} for the weekly incident rates
and \{1,1.1,1.5,2\} for the \(R_t\)'s. Anyway, these cuts could be
alternatively defined as a function of sample percentiles, for example,
if this was found more convenient. Colors in Figure 4 have been chosen
in order to represent different levels of increasing risk (higher as we
move towards the upper-right side of the plot), varying from greens for
the municipalities with lower rates and \(R_t\)'s and reds for the
municipalities with high values for these two indicators. This plot
shows how the highest rate corresponds to a municipality with around 800
incident cases per 100000 people and the highest \(R_t\) is above 2.5
for one municipality.

The right-hand side of Figure 4 shows the geographical mapping of the
risks shown in the left-hand side plot of this same figure.
Municipalities are colored in the right plot with the color (risk level)
corresponding to the region where they lay in the left side plot.
Interestingly, this choropleth map shows an evident spatial pattern,
pointing out several regions of particularly high or low risks. This
spatial pattern arises as a consequence of the spatio-temporal modelling
strategy followed, since otherwise municipalities (mainly those less
populated) would mostly show a noisy geographical pattern. This map
could allow the epidemiological CV authority to focus its efforts on
those municipalities that most require it.

\hypertarget{conclussions}{%
\section{Conclussions}\label{conclussions}}

This paper introduces small areas spatio-temporal analyses for Covid-19
epidemiological surveillance. Although the model introduced has been
applied to Spanish data, in our opinion the methods developed could be
perfectly applied to any other region with similar available data. In
order to run our model only population and daily incident cases data are
required, which does not seem a strong requirement.

We would like to mention an epidemiological limitation of the analysis
we have carried out in this paper. It is widely known that Covid-19
incidence data for both epidemic waves in our case study, at least for
Spain, are not comparable. The lack of diagnostic tests of the disease
during the first pandemic wave, for example, makes the available data
for that period to underestimate the height of that wave, making both
waves uncomparable in absolute terms. Nevertheless, the joint analysis
of all the epidemic period of the disease (both waves) has allowed
performing an overall analysis of the spatial distribution of the
disease during all that period. In principle, the spatial distribution
of the disease would not depend on the lack of diagnostic tests of the
disease since this was something generalised for all CV and CyL. Anyway,
direct comparisons of the strengths of the first and second epidemic
waves should be, and have been, avoided.

One interesting aspect of our proposal is its computational performance.
The dimensions of the data sets analysed for this paper are quite large,
with \(120286(=223\cdot 542)\) and \(55081(=223\cdot 247)\)
spatio-temporally correlated observations. Moreover, we have used a
heteroscedastic spatio-temporal dependence structure for our analyses,
which makes the underlying process more flexible but could also pose
additional computational problems that we have been able to deal with.
This has been possible due to the dimensionality reduction achieved by
the use of the spline basis, which has reduced the number of spatial
patterns to be fitted from 223 to just 17. This makes this fit of this
model substantially faster than spatio-temporal models with one spatial
process per time unit (Martinez-Beneito, López-Quílez, and
Botella-Rocamora 2008). Nevertheless, the computing times required for
fitting both data sets have been high (16.2 and 7.9 hours),
respectively. Possibly, these computing times could be improved by using
other MCMC sampling engines, such as Nimble or STAN. Nevertheless, for
the usual case of being surveillance the main purpose of the study, in
particular surveillance of the latest day of analysis, we would not need
to consider such a long period of study. Most of the functions of
natural B-spline bases have compact support so, for the latest day of
study, increasing the period of analyses does not yield any improvement.
The first splines in the basis will not take any effect on the final
part of the period of study. In this case, reducing the days of analysis
to the latest 56 days would reduce the spline basis to just 5 functions,
reducing accordingly the computing time (for CV this reduces the
computing time to just 1.16 hours), and producing the same fit for the
latest day that if we considered a higher number of days for the
analysis.

Most current Covid-19 epidemiological surveillance systems use incident
rates for the last 7 or even 14 days. Using 14 days rates may seem
justified as a proxy of disease prevalence, assuming that the disease
(or its contagious period) lasts in average around 14 days to vanish.
Similarly, 7 days rates seem also a reasonable option for removing the
weekly cyclic character that data could supposedly show. Moreover, these
two rates consider several days, increasing therefore the number of
observed cases used to compute each rate, what could alleviate the small
areas problems that data could show and enable to calculate \(R_t\)
estimates. Nevertheless, despite of these appealing features,
epidemiologic surveillance with 7 or 14 days rates does not seem the
best option. The aggregation of several days into intervals makes the
corresponding rates to average the information of those time intervals,
reducing therefore the novelty that surveillance epidemiological systems
should supposedly have. Effective surveillance systems need data as
actual and updated as possible and that feature is lost with the
temporal aggregation of data. In this sense our model allows focusing
our attention on the latest day of analysis, making Covid-19
surveillance more effective.

Undoubtedly, spatio-temporal analyses of the Covid-19 pandemics could
provide important insight on the dynamics of the disease. As
illustrated, these analyses have shown the effect of touristic activity
on the diffusion and spatial distribution of the disease. Moreover, we
have developed surveillance tools that allow to focus epidemiologists'
attention wherever is required in order to contain the advance of the
disease. Nevertheless, these analyses are useful only if small areas are
used to perform statistical studies. If large areas of analysis were
used instead, results and conclusions would be much less detailed and
powerful, what would reduce the impact of those studies. Therefore, the
development of small areas analysis tools seems necessary in order to
increase the value of spatial epidemiological studies against Covid-19.
This work is just a step in that direction.

\hypertarget{bibliography}{%
\section*{Bibliography}\label{bibliography}}
\addcontentsline{toc}{section}{Bibliography}

\hypertarget{refs}{}
\begin{cslreferences}
\leavevmode\hypertarget{ref-Besag.York.ea1991}{}%
Besag, Julian, Jeremy York, and Annie Mollié. 1991. ``Bayesian Image
Restoration, with Two Applications in Spatial Statistics.'' \emph{Annals
of the Institute of Statistical Mathemathics} 43: 1--21.
\url{https://doi.org/10.1007/BF00116466}.

\leavevmode\hypertarget{ref-CNISCI2020}{}%
Centro Nacional de Epidemiología. Insituto de Salud Carlos III. 2020.
November 1, 2020. \url{https://cnecovid.isciii.es/covid19/}.

\leavevmode\hypertarget{ref-ConsejoInterterritorial2020}{}%
Consejo Interterritorial, Sistema Nacional de Salud. 2020. ``Coordinated
Response Actions for the Transmission Control of Covid-19 {[}in
Spanish{]}.'' 2020.
\url{https://www.lamoncloa.gob.es/serviciosdeprensa/notasprensa/sanidad14/Documents/2020/221020_ActuacionesrespuestaCOVID.pdf}.

\leavevmode\hypertarget{ref-Cori2020}{}%
Cori, Anne. 2020. \emph{EpiEstim: Estimate Time Varying Reproduction
Numbers from Epidemic Curves}.
\url{https://CRAN.R-project.org/package=EpiEstim}.

\leavevmode\hypertarget{ref-Cori.Ferguson.ea2013}{}%
Cori, Anne, Neil M. Ferguson, Christophe Fraser, and Simon Cauchemez.
2013. ``A New Framework and Software to Estimate Time-Varying
Reproduction Numbers During Epidemics.'' \emph{American Journal of
Epidemiology} 178 (9): 1505--12.
\url{https://doi.org/10.1093/aje/kwt133}.

\leavevmode\hypertarget{ref-Fang.Nie.ea2020}{}%
Fang, Yaqing, Yiting Nie, and Marshare Penny. 2020. ``Transmission
Dynamics of the COVID-19 Outbreak and Effectiveness of Government
Interventions: A Data-Driven Analysis.'' \emph{Journal of Medical
Virology} 92 (6): 645--59. \url{https://doi.org/10.1002/jmv.25750}.

\leavevmode\hypertarget{ref-Gelman.Carlin.ea2014}{}%
Gelman, Andrew, Jon B Carlin, Hal S Stern, David B. Dunson, Aki Vehtari,
and Donald B Rubin. 2014. \emph{Bayesian Data Analysis}. 3rd ed. Boca
Raton: Chapman \& Hall/CRC.

\leavevmode\hypertarget{ref-Han.Tan.ea2020}{}%
Han, Emeline, Melisa Mei Jin Tan, Eva Turk, Devi Sridhar, Gabriel M
Leung, Kenji Shibuya, Nima Asgari, et al. 2020. ``Lessons Learnt from
Easing COVID-19 Restrictions: An Analysis of Countries and Regions in
Asia Pacific and Europe.'' \emph{The Lancet}, September.
\url{https://doi.org/10.1016/s0140-6736(20)32007-9}.

\leavevmode\hypertarget{ref-Lawson2018}{}%
Lawson, Andrew B. 2018. \emph{Bayesian Disease Mapping: Hierarchical
Modeling in Spatial Epidemiology (3rd Edition)}. CRC Press.

\leavevmode\hypertarget{ref-Leroux.Lei.ea1999}{}%
Leroux, Brian G., Xingye Lei, and Norman Breslow. 1999. ``Estimation of
Disease Rates in Small Areas: A New Mixed Model for Spatial
Dependence.'' In \emph{Statistical Models in Epidemiology, the
Environment and Clinical Trials}, edited by M E Halloran and D Berry.
Berlin Heidelberg New York: Springer.
\url{https://doi.org/10.1007/978-1-4612-1284-3_4}.

\leavevmode\hypertarget{ref-LopezCodina2020}{}%
López Codina, Daniel. 2020. ``Tools for Early Detection and Risk
Assessment.'' In \emph{COVID-19 \& Response Strategy}, edited by
ISGlobal.
\url{https://www.isglobal.org/en/-/-como-hacer-frente-a-los-nuevos-brotes-de-la-covid-19-}.

\leavevmode\hypertarget{ref-MartinezBeneito.BotellaRocamora2019}{}%
Martinez-Beneito, Miguel A., and Paloma Botella-Rocamora. 2019.
\emph{Disease Mapping from Foundations to Multidimensional Modelling}.

\leavevmode\hypertarget{ref-Martinez-Beneito.LopezQuilez.ea2008}{}%
Martinez-Beneito, Miguel A., Antonio López-Quílez, and Paloma
Botella-Rocamora. 2008. ``An Autoregressive Approach to Spatio-Temporal
Disease Mapping.'' \emph{Statistics in Medicine} 27: 2874--89.

\leavevmode\hypertarget{ref-Milligan.Barrett2014}{}%
Milligan, Gregg N., and Alan D. T. Barrett. 2014. \emph{Vaccinology: An
Essential Guide}. John Wiley \& Sons.
\url{https://www.ebook.de/de/product/23507913/vaccinology.html}.

\leavevmode\hypertarget{ref-Nishiura.Chowell2009}{}%
Nishiura, Hiroshi, and Gerardo Chowell. 2009. ``The Effective
Reproduction Number as a Prelude to Statistical Estimation of
Time-Dependent Epidemic Trends.'' In \emph{Mathematical and Statistical
Estimation Approaches in Epidemiology}, 103--21. Springer Netherlands.
\url{https://doi.org/10.1007/978-90-481-2313-1_5}.

\leavevmode\hypertarget{ref-Nishiura.Linton.ea2020}{}%
Nishiura, Hiroshi, Natalie M. Linton, and Andrei R. Akhmetzhanov. 2020.
``Serial Interval of Novel Coronavirus (COVID-19) Infections.''
\emph{International Journal of Infectious Diseases} 93 (April): 284--86.
\url{https://doi.org/10.1016/j.ijid.2020.02.060}.

\leavevmode\hypertarget{ref-Rao2003}{}%
Rao, J N K. 2003. \emph{Small Area Estimation}. John Wiley \& Sons.

\leavevmode\hypertarget{ref-Rue.Held2005}{}%
Rue, Havard, and Leonhard Held. 2005. \emph{Gaussian Markov Random
Fields: Theory \& Applications}. Chapman \& Hall/CRC.

\leavevmode\hypertarget{ref-Tebe.Valls.ea2020}{}%
Tebé, Cristian, Joan Valls, Pau Satorra, and Aurelio Tobías. 2020.
``COVID19-World: A Shiny Application to Perform Comprehensive
Country-Specific Data Visualization for SARS-CoV-2 Epidemic.'' \emph{BMC
Medical Research Methodology} 20 (1).
\url{https://doi.org/10.1186/s12874-020-01121-9}.

\leavevmode\hypertarget{ref-Torres-Aviles.Martinez-Beneito2015}{}%
Torres-Avilés, Francisco, and Miguel A. Martinez-Beneito. 2015.
``STANOVA: A Smooth-ANOVA-Based Model for Spatio-Temporal Disease
Mapping.'' \emph{Stochastic Environmental Research and Risk Assessment}
29 (1): 131--41. \url{https://doi.org/10.1007/s00477-014-0888-1}.

\leavevmode\hypertarget{ref-Vergara.MartinezBeneito2020a}{}%
Vergara-Hernández, Carlos, and Miguel A. Martinez-Beneito. 2020.
``Pbugs.'' September 29, 2020. \url{https://github.com/fisabio/pbugs}.

\leavevmode\hypertarget{ref-Wilasang.Sararat.ea2020}{}%
Wilasang, Chaiwat, Chayanin Sararat, Natcha C Jitsuk, Noppamas Yolai,
Panithee Thammawijaya, Prasert Auewarakul, and Charin Modchang. 2020.
``Reduction in Effective Reproduction Number of COVID-19 Is Higher in
Countries Employing Active Case Detection with Prompt Isolation.''
\emph{Journal of Travel Medicine} 27 (5).
\url{https://doi.org/10.1093/jtm/taaa095}.
\end{cslreferences}

\bibliographystyle{unsrt}
\bibliography{BibliografiaUTF8.bib}

\end{document}